\documentstyle[12pt]{article} 
\begin{document}

\def\be{\begin{equation}}
\def\ee{\end{equation}}

\begin{flushright}
HRI-RECAPP-2011-005
\end{flushright}

\begin{center}
{\large\bf Matter-gravity interaction in a multiply warped braneworld}\\[10mm]

Biswarup Mukhopadhyaya\footnote{E-mail: biswarup@hri.res.in}\\
{\em Regional Centre for Accelerator-based Particle Physics\\
Harish-Chandra Research Institute\\
Chhatnag Road, Jhusi, Allahabad - 211 019, India}

Somasri Sen \footnote{E-mail: somasri.ctp@jmi.ac.in}\\
{\em Centre for Theoretical Physics\\
Jamia Millia University\\
New Delhi 110 025, India}

Soumitra SenGupta \footnote{E-mail: tpssg@iacs.res.in} \\
{\em Department of Theoretical Physics\\
 Indian Association for the 
Cultivation of Science\\
Calcutta - 700 032, India}\\

\vspace{0.1cm}
\end{center}

\begin{abstract}
The role of a bulk graviton in predicting the signature of extra dimensions through collider-based experiments is explored in the context of 
a multiply warped spacetime. In particular it is shown that in a doubly warped braneworld model, the presence of the sixth dimension,   
results in enhanced concentration of graviton Kaluza Klein (KK) modes compared to that obtained in the usual 5-dimensional Randall-Sundrum  model.
Also, the couplings of these massive graviton KK modes with the matter fields on the visible brane turn out to be appreciably
larger than that in the corresponding 5-dimensional model. The significance of these results are discussed in the context of KK graviton search at the Large Hadron Collider (LHC).   
\end{abstract}

\vskip 1 true cm

\newpage
\section{Introduction}
The possibility that spacetime may have more than four dimension is often seriously considered in high energy physics. Such scenarios are also strongly supported by string theory. Several new ideas in this direction have evolved to explore various implications of extra spatial dimensions in the context of particle 
physics and cosmology. In fact, according to some recent proposals the observed weakness  of gravity may be due to the fact that we live on a 3-brane embedded in space with large extra dimension. 
One of the most prominent theories developed in the context of braneworld 
models is due to Randall and Sundrum (RS)\cite{rs1}. The proposal was originally aimed at bridging the large hierarchy between the electroweak $(m_W\sim 10^2GeV)$ and the Planck $(M_P\sim 10^{18}GeV)$ scales through the existence of an extra spatial dimension with warped geometry. The Randall-Sundrum model assumes $AdS_5$ spacetime with the extra spatial dimension orbifolded as $S_1 /Z_2$ bounded by two 4D Minkowski branes called IR(TeV) and UV (Planck) branes.    
The essential input of this theory is that gravity propagates in the bulk spacetime, while the standard model (SM) fields lie on a brane. 
The curvature in 5D induces a warped geometry on the brane which redshifts the scale of order $M_P$ at UV brane to the scale of 
order $m_W$ at IR brane. A lot of work exploring various aspects of this model has been carried out in the last few years. This includes its 
role in resolving the hierarchy problem\cite{rs1,besan}, localization of various types of fields on the brane\cite{bajc}, particle phenomenology in context 
of braneworld \cite{besan}, various cosmological consequences\cite{cosmo} and so on. A scheme to achieve stability issue of this model was proposed by Goldberger and Wise (GW)\cite{gw} and the effects of other bulk fields
like gauge field or higher form fields been studied in several
works~\cite{ssg,ferreira,bulk,davod1,davod2,our, cvetic}.
Though a firm foundation for these models still remain an open issue,it has been widely recognised that one of the key signature for extra dimension can be obtained from collider physics by observing various KK resonances.       

As a natural extension to the RS scenario, models with more than one warped extra dimensions have been proposed\cite{6dmodel1,6dmodel2}. Most of these models consider several independent $S_1/Z_2$ orbifolded dimension along with $M_4$. It is apparent from different considerations that the radion stabilization problem \cite{leblond,kogan} and the presence of negative tension brane are artifacts of 5D spacetime and are avoidable in a  higher dimensional generalization. 

In yet another interesting scenario\cite{dcssg}, a warped compact dimension get further warped by a series of successive warping leading to {\em multiple warping of the spacetime} with various p-branes sitting at the different orbifold fixed points satisfying appropriate boundary conditions. Various lower dimensional branes along with the standard model 3-brane exist at the intersection edges of the higher dimensional branes. The resulting geometry of the D dimensional spacetime is $ M^{1,D-1}\rightarrow \{[M^{1,3}\times S^1/Z_2]\times S^1/Z_2\}\times....$ with $D-4$ warped spacelike dimensions. It has been argued that this multiply warped spacetime gives rise to interesting phenomenology and offers a possible explanation of small mass splitting among the standard model fermions\cite{dcssg, rsj1}.  One more interesting feature of this model is the bulk coordinate dependence of the higher dimensional brane tensions unlike 5D RS model
where brane tensions are constant. In this work we want to explore for such multiply warped spacetime, how gravity appears on the 3-brane where visible matter fields reside

As in 5D RS the essential input in this multiply warped scenario is that gravity propagates in the bulk spacetime, while the standard model (SM) fields are assumed to lie on a 3-brane. This type of description gets a strong support from string theory where the SM fields arise as open string excitations 
whose ends are fixed on the brane while the graviton being a closed string excitation can propagate in the bulk space-time\cite{pol}. 
In 5D RS the massless graviton mode has a coupling $\sim 1/M_P$ with all matter on the brane , while the massive KK modes have enhanced coupling
through the warp factor. Thus, the massless graviton mode accounts for the presence of gravity in our universe, while the massive modes raise 
hopes for possible new signals of extra warped dimension in accelerator experiments\cite{davod1}.
In this work we investigate the implications of the KK tower of gravitons in multiply warped spacetime, specifically a doubly warped spacetime. We intend to find the mass spectrum and their interaction with matter on the visible brane.

The present article is organized as follows. In the next section, we briefly discuss the multiply warped six-dimensional model. In the third section we focus on the graviton and find out its equation of motion
and the Kaluza Klein modes. In section 4, we consider the interaction of different KK modes of graviton with matter on visible brane and 
find their coupling strengths. We summarise and conclude in section 5.

\section{Multiply warped brane world model in 6D}    

Our working model is a doubly warped compactified six-dimensional spacetime 
with a $Z_2$ orbifolding in each of the extra dimensions. The manifold under consideration is $M^{1,5}\rightarrow[M^{1,3}\times S^1/Z_2]\times S^1/Z_2$. 
The metric representing such a spacetime is given by
\be
\label{metric}
ds^2_6= b^2(z)[a^2(y)\eta_{\mu\nu}dx^{\mu}dx^{\nu}+{R_y}^2dy^2]+{r_z}^2dz^2
\ee  
We use Greek indices for the noncompactified directions while the orbifolded compact directions are represented by coordinates $y$ and $z$. $R_y$ and $r_z$ are the two moduli corresponding to $y$ and $z$ directions and $a(y)$ and $b(z)$ are the corresponding warp 
factors. $\eta_{\mu\nu}$ is the Minkowski metric $(-,+,+,+)$. Following the orbifold requirement, four 3-branes are located at four orbifold 
fixed points given by $y=(0,\pi)$ and $z=(0,\pi)$.

The total bulk-brane action for the six dimensional space time is,
\begin{eqnarray}
{\cal{S}}&=&{\cal{S}}_6+{\cal{S}}_5+{\cal{S}}_4\nonumber\\
{\cal{S}}_6&=&\int d^4x dy dz \sqrt{-g_6}(R_6-\Lambda)\nonumber\\
{\cal{S}}_5&=&\int d^4x dy dz [V_1\delta(y)+V_2\delta(y-\pi)]\\
&+&\int d^4x dy dz [V_3\delta(z)+V_4\delta(z-\pi)]\nonumber\\
{\cal{S}}_4&=&\int d^4x dy dz \sqrt{-g_{vis}}[{\cal{L}}-\hat{V}]\nonumber
\end{eqnarray}
where $\Lambda$ is the bulk cosmological constant which is necessarily negative. In general, 
the brane tensions are $V_{1,2}=V_{1,2}(z)$ and  $V_{3,4}=V_{3,4}(y)$. $S_4$ represents the 3-branes located at
$(y, z) = (0, 0), (0, \pi), (\pi, 0), (\pi, \pi)$. 
Einstein field equations for metric(\ref{metric}) in the above action leads to the following solutions for the  warp factors
\begin{eqnarray}
a(y)= e^{-\rho|y|} &\hspace{2cm}& \rho=\frac{R_y k}{r_z\cosh{k\pi}}\\
b(z)=\frac{\cosh{(kz)}}{\cosh{(k\pi)}} &\hspace{2cm}& k=r_z\sqrt{\frac{-\Lambda}{M^4}}\nonumber
\end{eqnarray}
It should be noted that the solutions are $Z_2$ symmetric about the $y$
and $z$ directions. The brane tensions are obtained by considering the 
boundary terms. The brane tensions at the two boundaries $y = 0$ and $y =\pi$ 
are given by
\be
\label{yten}
V_1(z)=-V_2(z)=8M^2\sqrt{\frac{-\Lambda}{10}} sech{(kz)}
\ee
 The two 4-branes sitting at $y = 0$ and $y = \pi$ have equal and opposite tensions but, unlike in the 5D RS model, the tensions are z-dependent. Similarly, the boundary condition for the infinitesimal interval across $z = 0$ and
$z = \pi$ leads to
\be
\label{zten}
V_3(y)=0,\hspace{1cm} V_4(y)=-\frac{8M^4k}{r_z}\tanh{(k\pi)}
\ee
In this case the brane tensions are constants similar to the original RS model. The fact that $g_{yy}$ is a 
non-trivial function of $z$ leads to the two hypersurfaces for the $y$ orbifolding to have z-dependent 
energy density. If there exists no other brane with an energy scale lower than ours, we must identify the SM brane with the one at $y = \pi, z = 0$.

In this model the solution of the hierarchy problem (i.e. the mass rescaling due to warping) demands that unless there is a large hierarchy between 
the moduli 
$r_z$ and $R_y$ , either of $\rho$ and $k$ must be small implying large warping
in one direction and small in the other. This particular feature of this model 
is revealed from the relation,
$\rho =\frac{R_y k}{r_z cosh k\pi}$,
which implies that for $R_y \sim r_z$ a large hierarchy along y-direction (a situation very close in spirit with RS) and a relatively small 
$k\sim {\cal{O}}(1)$ warping in the z direction. In summary, we are dealing with a braneworld which is doubly warped,  the warping is large along 
one direction and small in the other. We would now address the nature of Kaluza Klein modes of graviton in such a six dimensional braneworld.

\section{The Graviton KK modes}

\subsection{Massive modes}

To calculate the mass spectrum  and couplings of the graviton KK modes in effective 4 dimensional theory on the 3-brane, we first parameterize the tensor fluctuations $h_{\alpha\beta}$ by taking a linear expansion of the flat metric about its Minkowski value $\hat{G_{\alpha\beta}}=b^2(z)a^2(y)[\eta_{\alpha\beta}+k^\star h_{\alpha\beta}]$ where $k^\star$ is the expansion parameter  is ${\cal{O}}(\frac{1}{M^2})$. 


With the gauge choice $\partial^\alpha h_{\alpha\beta}=h^\alpha_\alpha=0$ the 6D Einstein Hilbert action in the linearised gravity limit appears as
\be 
{\cal{S}}=\frac{1}{4}\int d^4 x \int dy \int dz \sqrt{-g_6}[\partial^M {h}_{\alpha\beta}\partial_M {h}_{\alpha\beta}] 
\ee
In order to obtain the mass spectrum of the tensor fluctuations we expand the graviton field $h_{\alpha\beta}$ in a KK tower 
\be
\label{hdeco}
{h}_{\alpha\beta}(x^\mu,y,z)=\sum_n\sum_p\frac{1}{\sqrt{R_yr_z}}{h}^{np}_{\alpha\beta}(x^\mu)\xi_n(y)\chi_p(z)
\ee
where the ${h}^{np}_{\alpha\beta}$ correspond to the KK modes of the graviton on the background of Minkowski space on 3-brane. Upon compactification the effective 4D graviton action 
\be
{\cal{S}}=\frac{1}{4}\int d^4 x \sum_{np}[\eta^{\mu\nu}\partial_\mu{h}^{np}_{\alpha\beta}\partial_\nu{h}^{np}_{\alpha\beta}+m^2_{np}h^{np}_{\alpha\beta}]
\ee
is obtained provided $\xi$ and $\chi$ satisfies the following normalization conditions 
\be
\label{axi}\int_{-\pi}^{\pi} a^2\xi_{n1}(y)\xi_{n2}(y) dy =\delta_{n1n2}
\ee
\be
\label{bchi}\int_{-\pi}^{\pi} b^3\chi_{p1}(z)\chi_{p2}(z) dz =\delta_{p1p2}
\ee
respectively with their respective eigenvalue equations  
\be
\label{xi}
-\frac{1}{R_y^2}\partial_y(a^4\partial_y\xi_n) + a^4 m_p^2\xi_n=m_{np}^2 a^4\xi_n
\ee
\be
\label{chi}
\frac{1}{r_z^2}\partial_z(b^5\partial_z\chi_p) + b^3 m_p^2\chi_p=0\label{chi}
\ee
As in usual Kaluza Klein compactification, the fluctuations in the bulk ${h}_{\alpha\beta}(x^\mu,y,z)$ appear to a four dimensional observer as 
an infinite tower of tensor modes ${h}^{np}_{\alpha\beta}$ with mass $m_{np}$. Point to notice here is that the KK mass term carries two indices 
because of two compact warped extra dimensions. This implies that the usual five dimensional massive tower further splits into sub tower due to 
the additional warped dimension. These extra modes naturally will have their contributions in the collider experiments and are expected to 
produce enhanced signature for extra dimension. 

The KK  tower corresponding to the extra dimension $z$ is given by $m_p$ whereas the full four dimensional KK mass tower is represented by $m_{np}$. 
It may be noted  that the mass tower $m_p$ due to $\chi_p(z)$ enters in the equation for $\xi_n(y)$ to determine $m_{np}$. So we have to determine the mass 
tower $m_p$ first and then use it equation (\ref{xi}) to achieve the KK mass spectrum in the visible 3-brane. 

Considering the allowed domain for $k$ we approximate the warp factor $b(z)\sim e^{-k(\pi-z)}=e^{-k \bar{z}}$. Redefining $z$ in terms of a 
new variable $z_p=\frac{m_p}{k^\prime}e^{k\bar{z}}$ where $k^\prime=\frac{k}{r_z}$ and $\chi_p(z)$ in terms of a new 
function $\bar{\chi_p}$, where $\bar{\chi_p}=e^{-\frac{5}{2}k\bar z}\chi_p(z)$,  equation(\ref{chi}) takes the form of a Bessel Differential equation of order $5/2$
\be
z_p^2\frac{\partial^2\bar\chi_p}{\partial z_p^2}+z_p\frac{\partial\bar\chi_p}{\partial z_p}+(z_p^2-\frac{25}{4})\chi_p=0
\ee
Evidently the solution for $\chi_p$ is then given by Bessel function of order $\frac{5}{2}$
\be
\label{solnchi}
\chi_p(\bar{z})=\frac{e^{\frac{5}{2}k\bar{z}}}{N_p}[J_{5/2}(\frac{m_p}{k^\prime}e^{k\bar{z}})+\alpha_p Y_{5/2}(\frac{m_p}{k^\prime}e^{k\bar{z}})]
\ee
where $J_{5/2}$ and $Y_{5/2}$ are Bessel and Neumann functions of order $5/2$, $N_p$ is normalization constant and $\alpha_p$ is an arbitrary 
constant. $\alpha_p$ and $m_p$ are determined by the continuity conditions of the first derivative of $\chi_p$ at the orbifold fixed points $z=0 \& \pi$, 
which are dictated by the self adjointness condition of the left hand side of equation (\ref{chi}). The continuity conditions 
implies $\alpha_p<<1$ and the spectrum for $m_p$ is to be obtained from the roots of 
\be\label{chimass}
J_{3/2}(x_p)=0, \hspace{3cm} \mbox{where}~~~~ x_p=\frac{m_p}{k^\prime}e^{k\pi}  
\ee
The normalization condition (\ref{bchi}) determine the normalization constant 
\be
\label{Np}
N_p=\frac{\sqrt{2}e^{k\pi}}{\sqrt{k^\prime}r_z}B_p
\ee
where $B_p^2=\int^1_0sJ_{5/2}^2(x_p s)ds$ with $s=e^{-kz}$. Once we find  the z dependent part of the KK modes, we solve for the y dependent part of the modes to arrive at the full spectrum 
of the KK modes on the visible 3-brane. Adopting a similar technique, equation(\ref{xi}) can be recast in the form 
of a Bessel differential equation of order $\nu$, where $\nu= \sqrt{4+\frac{m_p^2}{k^{\prime 2}}}$. Hence, the solution for $\xi_n(y)$ can be  expressed 
in terms of Bessel and Neumann function of order $\nu$
\be 
\label{solnxi}
\xi_n(y)=\frac{e^{2c|y|}}{N_n}[J_{\nu}(\frac{m_{np}}{k^\prime}e^{\rho|y|})+\alpha_{np} Y_{\nu}(\frac{m_{np}}{k^\prime}e^{\rho|y|})]
\ee
$N_n$ and $\alpha_{np}$ are the two constants. Once again the condition of self adjointness leads us to the mass spectrum for $m_{np}$ through the 
following transcendental equation 
\be
\label{ximass}
2J_\nu(x_{np})+x_{np}J_\nu^\prime(x_{np})=0 ~~~~\mbox{where}~~~~x_{np}=\frac{m_{np}}{k^\prime}e^{c\pi}
\ee
In the five dimensional case the mass spectrum of the graviton KK excitations are determined from the roots of the first order 
Bessel function $J_1$. However, in the six-dimension we have several orders of the Bessel function representing the KK modes 
because the order $\nu$ can take up different values for different values of $m_p$. Therefore for each mass 
splitting due to z-compactifiction we  obtain a spectrum of KK modes. Hence, we obtain extra splitting in the spectrum over the usual 
5-dimensional scenario. In Table 1 we show explicitly the masses of the KK modes for three different values of $k$.

\begin{table}[h]
\begin{tabular}{|c|c|c|c|c|c|c|c|c|c|c|c|c|}
   \hline 
 $m_{np}$ &  \multicolumn{4}{|c|}{k= 0.1}  & \multicolumn{4}{|c|}{k=0.05} & \multicolumn{4}{|c|}{k=0.3}\\ 
 \cline{2-13}
 & p=0 & p=1 & p=2 & p=3 &  p=0  & p=1 & p=2 & p=3 & p=0 & p=1 & p=2 &p=3 \\ \hline
 n=1& $m_{10}$ & $m_{11}$  & $m_{12}$ &$m_{13}$&$m_{10}$ &$m_{11}$  & $m_{12}$&$m_{13}$& $m_{10}$& $m_{11}$&$m_{12}$&$m_{13}$ \\
& 2.001 & 3.24 & 4.56 & 5.94&1.93 & 3.43 & 4.97 & 6.49& 2.83 & 3.47 & 4.36 & 5.33\\           
 \hline
 n=2 & $m_{20}$ & $m_{21}$  & $m_{22}$ &$m_{23}$&$m_{20}$ &$m_{21}$  & $m_{22}$&$m_{23}$& $m_{20}$& $m_{21}$&$m_{22}$&$m_{23}$ \\
 & 3.68 & 5.01 & 6.46 &7.91 &3.55 & 5.61 & 6.81 & 8.44 & 5.18 & 5.87 & 6.83 & 7.88\\          
\hline
  n=3&$m_{30}$ & $m_{31}$  & $m_{32}$ &$m_{33}$&$m_{30}$ &$m_{31}$  & $m_{32}$&$m_{33}$ &$m_{30}$& $m_{31}$&$m_{32}$&$m_{33}$ \\ 
& 5.34 & 6.71 & 8.23 & 9.74 & 5.45 &6.81  & 8.54  & 10.23 & 7.51 & 8.22 & 9.22 &10.31\\           
\hline
 
  \end{tabular}


 \caption{The KK mode masses $m_{np}$ (in TeV) for different values of $n$ and $p$, for $k^{'}/M=1$. }

\end{table}
 
 Once again, like the 5-dimensional case, the graviton KK
mode masses are suppressed by the warp factor.  We find that the light
KK modes have masses in the range of TeV. In this case, however, a much larger number of KK modes of mass $\sim$TeV appear in 
comparison to the 5-dimensional counterpart. Within a range of 10 TeV, there are only 6 modes in 5-D RS model, whereas in the 
6D doubly warped spacetime we have a lot (Table 1) more than that. 
These modes are expected to produce additional contributions to 
various processes involving KK modes of gravitons in the forthcoming collider experiments at TeV scale. 

Like $N_p$, $N_n$ is also determined from the normalization condition for $\xi_n(y)$ given by equation(\ref{axi}) as
\be
\label{Nn}
N_n=\frac{\sqrt{2~cosh(k\pi)}~e^{\rho\pi}}{\sqrt{k^\prime R_y}}A_n
\ee

where $A_n^2=\int^1_0rJ_{\nu}^2(x_n r)dr$ with  $r=e^{\rho(y-\pi)}$.

\subsection{Massless modes of Graviton}

So far we have checked the higher KK modes of the tensor fluctuation. Now we find the status of the massless mode of the tensor fluctuation.
As the massless mode accounts for the observed gravity in our universe, it is important to find the lowest lying massless mode of 
the KK tower which corresponds to $n=0,p=0$. To find the the solution for the massless mode we solve equations (\ref{chi}) 
and (\ref{xi}) with $m_{p}=0$ and $m_{np}=0$ respectively. The solutions for $\chi_0$ and $\xi_0$ are
\be
\label{zero} 
\chi_0=\sqrt{\frac{3}{2}k^{\prime}r_z} ~~~~~\mbox{and} ~~~~~~\xi_0=\sqrt{\rho}
\ee

Here we would like to mention that for $p = 0$ , $n$ can have both zero and 
non-zero values. Among these  $n = 0, p = 0$ corresponds  to our massless 
graviton and rest all are different massive KK modes of graviton. But for $p\neq 0$, $n$ can have only nonzero values. This can be 
understood from the fact that $p\neq 0$ implies $m_p\neq 0$ 
which practically act as a bulk mass for y direction (index n) as is evident from equation (\ref{xi}). If a field has a bulk mass then 
there will be no massless KK mode. So for $p\neq 0$, $n$ ranges from $1$ to $\infty$. Thus no mode corresponding to $n=0, p\neq 0$ exists.

\section{Interaction on the TeV brane}

The final solution for tensor fluctuations that appear on our visible brane can be obtained by substituting the solution for $\chi$ 
and $\xi$ in equation (\ref{hdeco}) at $y=\pi$ and $z=0$
\begin{eqnarray}
{h}_{\alpha\beta}(x^\mu,y=\pi,z=0)=\sum_{n=0}^{\infty}\sum_{p=0}^{\infty}\frac{1}{\sqrt{R_y r_z}}{h}^{np}_{\alpha\beta}(x^\mu)\xi_n(\pi)\chi_p(0)\nonumber\\
=\frac{\sqrt{3}k^\prime}{\sqrt{2~cosh(k\pi)}}{h}^{00}_{\alpha\beta}(x^\mu)
+\sum_{n=1}^{\infty}\frac{k^\prime}{2A_n}\frac{\sqrt{3}e^{\rho\pi}}{\sqrt{cosh(k\pi)}}J_\nu(x_{n0}){h}^{n0}_{\alpha\beta}(x^\mu)\nonumber\\
+ \sum_{n=1}^{\infty}\sum_{p=1}^{\infty}\frac{k^\prime}{2A_nB_p} \frac{e^{(\rho+\frac{3}{2}k)\pi}}{\sqrt{cosh(k\pi)}}J_\nu(x_{np})J_{5/2}(x_p){h}^{np}_{\alpha\beta}(x^\mu)
\end{eqnarray}
So the interaction Lagrangian in the effective 4D theory is
\be
\label{lag}
{\cal{L}}=\frac{1}{M^2}~T^{\alpha\beta}_{SM}(x^\mu)~{h}_{\alpha\beta}(x^\mu,y=\pi,z=0)
\ee
where $T^{\alpha\beta}_{SM}(x^\mu)$ is the energy momentum tensor of the standard model matter field on the visible brane. 
In terms of the solution expressed above, equation(\ref{lag}) takes the form
\begin{eqnarray}
{\cal{L}}=\frac{k^\prime}{M^2}~\frac{\sqrt{3}}{\sqrt{2~cosh (k\pi)}}~{h}^{00}_{\alpha\beta}(x^\mu)~T^{\alpha\beta}_{SM}
+\frac{k^\prime}{M^2}~\sum_{n=1}^{\infty}\frac{1}{2A_n}\frac{\sqrt{3}e^{\rho\pi}}{\sqrt{cosh(k\pi)}}J_\nu(x_{n0})~{h}^{n0}_{\alpha\beta}(x^\mu)~T^{\alpha\beta}_{SM}\nonumber\\
+\frac{k^{'}}{M^2}~\sum_{n=1}^{\infty}\sum_{p=1}^{\infty}\frac{1}{2A_nB_p} 
\frac{e^{(\rho+\frac{3}{2}k^\prime)\pi}}{\sqrt{cosh k\pi }}J_\nu(x_{np})J_{5/2}(x_p)~{h}^{np}_{\alpha\beta}(x^\mu)~T^{\alpha\beta}_{SM}
\end{eqnarray}
We see that the zero mode i.e the massless graviton mode 
couples with matter with the strength $1/M$ which is  of order similar to that of $1/M_p$. This is like the 5d case and 
is consistent with our observation. However the massive states are suppressed with different coupling strengths. 
Using the fact that $\sqrt{cosh(k\pi)}\sim 1$ for small $k$ and $e^{\rho\pi}=\frac{10^{16}}{cosh(k\pi)}$ (as demanded by the hierarchy problem) 
we have calculated the coupling strength for some low lying massive states as shown in the table below : 


\begin{table}[h]
\begin{tabular}{|c|c|c|c|c|c|c|c|c|c|}
   \hline 
${\cal{L}}$ &  \multicolumn{3}{|c|}{k= 0.1}  & \multicolumn{3}{|c|}{k=0.05} & \multicolumn{3}{|c|}{k=0.3}\\ 
\cline{2-10}
 & p=0 & p=1 & p=2 & p=0  & p=1 & p=2 & p=0 & p=1 & p=2  \\ \hline
 n=1 & 2.27 & 2.98  & 2.96 & 2.4 & 2.48 & 2.47 & 1.35 & 4.58 & 4.57\\           \hline
 n=2 & 2.27 & 2.97 & 2.97 & 2.4 & 2.49 & 2.48 & 1.35 & 4.59 & 4.59\\          
\hline

\end{tabular}


 \caption{The coupling strength of the interaction Lagrangian (in $k^{'}/M$ TeV$^{-1}$) for different values of $n$ and $p$. }

\end{table}

As is clear from Table 2,  the coupling strength of the interaction Lagrangian for massive states are of the order of weak scale. 
A point to be noted here is that the massive modes with $n=0$ and $p\neq 0$ are suppressed by 
Planck scale as in the z direction we have  little warping while the large warping exist in the y direction corresponding to the index $n$. 
However the states with both  $n \neq 0$ and $p\neq 0$  
have large coupling and the proliferation in number of such KK modes over the usual 5D RS model makes the signature of 
multiply warped models distinct and significant in respect to collider experiments. 

In the 5D RS model, the coupling of the massive graviton modes to matter fields is driven by
${\frac {\sqrt{c}}{\Lambda_\pi}}$, where $c=k/M_{Pl}$ and $\Lambda_\pi = M_{Pl} e^{-k r_c \pi}$. 
Here $M_{Pl}$ denotes the 4D Plank mass, M, the 5D Planck mass, and k is related to the bulk cosmological constant\cite{rs1}. 
According to some recent estimates, the first excited graviton state with mass upto 
about 3.8 TeV can be explored (mostly through $e^+ e^-$ and $\mu^+\mu^-$ invariant mass peaks)
at the Large Hadron Collider (LHC) with a centre-of-mass energy of 14 TeV and an integrated luminosity
of 100 $fb^{-1}$,  with c = 0.1 and $\Lambda_\pi$ = 10 TeV\cite{lhc} .  $c \le 0.1$ is 
motivated in the 5D model by the urge to achieve magnitudes of the brane tension similar to that In string-inspired scenarios, and
also with gauge coupling unification in view\cite{davod1}. The constraint, however, is based on somewhat simplifying
assumptions, non-consideration of non-perturbative effects etc. Thus the choice of $c$ upto unity
is difficult to rule out in a completely full-proof manner.

In the 6D scenario investigated by us, the interaction of massive graviton states to matter is
controlled by $\alpha k^{'}/M$ TeV$^{-1}$, where $\alpha$ is above 2 for the lowest-lying 
states (see Table 2),
and  {\em M  is the 6D Planck mass}. No constraint can be {\it a priori} imposed to disallow
$k'/M \simeq$ 1 in this scenario, as the brane tension here is not a constant but a function of $z$ . 
Moreover the 6D Planck mass $M$ can be tuned using the values of both the moduli along the two
warped directions following the relation given in \cite{dcssg}.
If that be the case, then the matter-graviton couplings for several lowest-lying states exceed 
that for the 5D case (for $c =$ 0.1,   $\Lambda_\pi$ = 10 TeV) by an approximate 
factor of 7 even for $k'/M \simeq$ 0.1.  Thus the graviton search limit goes up compared to
that of the 5D case mentioned above.  The situation is even more optimistic for a higher value
of $k'/M$. 

The above discussion brings out a rather interesting feature of the 6D RS scenario
in the context of LHC searches. The close proximity of the various graviton excited modes, 
as seen from Table 1,  makes it very likely to for them to be seen as a series of closely lying
resonances, thanks to the enhanced coupling pointed out in the previous paragraph.
This is in clear contrast to the 5D RS case,  where only the first excited graviton state
can be realistically expected to be seen, the others being somewhat higher up in mass.
One can easily see that this idea can be extended to incorporate even larger number of warped extra dimensions \cite{dcssg} where 
due to the presence of additional KK numbers more and more graviton KK modes appear within a specified energy range.
The corresponding matter-graviton couplings would also increase accordingly.
Thus the possibility  of  observing a  set of closely spaced 
massive graviton resonances at the LHC emerges as a rather remarkable prediction
of the scenario explored by us.

\section{Conclusion}
 This work investigates, in a multiply warped 6D spacetime, how gravity behaves on the visible brane. We have considered a doubly warped compactified six-dimensional spacetime with a $Z_2$ orbifolding in each of the extra dimensions. In such a manifold there are four 3-branes at the four orbifold fixed points. Among which standard model exist on the brane at $y=\pi, z=0$. The speciality of this model is the warping is large in one direction (namely, $y$) and small in the other direction $z$ where $k<1$. The other crucial feature which makes it distinct from RS is the brane tension, which are coordinate dependent in this model.

In this six dimensional spacetime gravity resides in the bulk. We calculate various Kaluza Klein modes as we compactify the tensor fluctuations. The bulk fluctuations appear to a four dimensional observer as an infinite tower of tensor modes as usual, but with masses which has two indices $m_{np}$. This happens as the usual five dimensional massive tower further splits into sub tower due to the additional warped dimension. The mass spectrum of the graviton KK excitations are determined from the Bessel function of order $\nu$, where $\nu=f(m_p)$, in contrast to the first order Bessel function in 5D. Therefore for each mass splitting due to z-compactifiction we obtain a spectrum of KK modes. This enhances the number of KK modes in the TeV range many times over the usual 5D case (refer to Table 1). 

 To investigate the role of graviton through collider-based experiments, we consider the interaction of various modes with other standard model fields. The massless mode separates from the tower and couples with the usual 4-dimensional strength $1/M_{Pl}$, like the 5 dimensional case. However, all the massive states in 6D are coupled by $\alpha k^{'}/M$ TeV$^{-1}$ ($\alpha$>2 for the lowest-lying states) in contrast to the coupling strength of $\sqrt{c}$TeV$^{-1}$ ($c\leq 1$) in 5D case. So, even for $k'/M \simeq$ 0.1, the matter-graviton couplings for several lowest-lying states in 6D is almost 7 times the 5D case. This  ratio gets even more optimistic for higher values of $k^{'}/M$. Thus the graviton search limit goes up, thereby enhancing the possibility of observing a set of closely spaced massive graviton resonances at the LHC.

\section{Acknowledgement}
The work of SS is financially supported by Department of Science and 
Technology, India under Fast Track Scheme (SR/FTP/PS-104/2010). SS also acknowledges the hospitality of Regional Centre for Accelerator-based Particle Physics,
Harish-Chandra Research Institute, where a part of this work is carried out. The work of BM was partially supported by funding available
from the Department of Atomic Energy, Government of India, for the
Regional Centre for Accelerator-based Particle Physics,
Harish-Chandra Research Institute.


\begin{thebibliography}{99}

\bibitem{rs1} L. Randall and R. Sundrum, Phys. Rev. Lett. {\bf 83}, 3370 (1999)

\bibitem{besan} M. Besancon, {\em Experimental introduction to
extra dimensions}, hep-th/0106165 ; 
Y. A. Kubyshin, {\em Models with
extra dimensions and their phenomenology} hep-ph/0111027
and references therein.

\bibitem{bajc} B. Bajc and G. Gabadadze, Phys. Lett. {\bf B 474}, 282
  (2000);  
S. L. Dubovsky, V. A. Rubakov and P. G. Tinyakov,
  Phys. Rev. {\bf D 62}, 105011 (2000)

\bibitem{cosmo} D. Langlois, gr-qc/0102007 (Proceedings of the
9th Marcel Grossmann meeting, July 2000 (Rome)); \newline  gr-qc/0205004 (Proceedings
of Journees Relativistes, Dublin 2001) and references therein
; \newline  E. Flannagan, S. H. Henry Tye and I. Wasserman,
Phys. Rev. {\bf D 62}, 024011 (2000);  \newline H. Stoica, S. H. Henry Tye
and I. Wasserman, Phys. Letts. {\bf B 482}, 205 (2000);  \newline J. S. Alcaniz
, astro-ph/0202492

\bibitem{gw} W.D. Goldberger and M. B. Wise, 
   Phys. Rev. Lett. {\bf 83} 4922 (1999); Phys. Lett. {\bf B475} 275 (2000)

\bibitem{ssg}  S. Das, A. Dey and S. SenGupta, 
Class.Quant.Grav.{\bf 23} L67 (2006); 
R. Maartens, Living Rev. Rel. {\bf 7} 7 (2004) and references therein;

\bibitem{ferreira} P.C. Ferreira and P.V. Moniz, hep-th/0601070; 
G.L. Alberghi and A. Tronconi, Phys. Rev. {\bf D73} 027702 (2006); 
A.A. Saharian and M.R. Setare, Phys. Lett. {\bf B552} 119 (2003).

\bibitem{bulk} W.D. Goldberger and M.B. Wise, 
Phys. Rev. {\bf D60} 107505 (1999);
S. Kachru, M.B. Schulz and E. Silverstein, Phys. Rev. {\bf D62} 045021 (2000); 
H.A. Chamblin and H.S. Reall, Nucl. Phys. {\bf B562} 133 (1999); 
C. Csaki, hep-ph/0404096; 
R. Neves, TSPU Vestnik {\bf 44N7} 94 (2004); 
E. Dudas and M. Quiros, Nucl. Phys. {\bf B721} 309 (2005). 

\bibitem{davod1}H. Davoudiasl, J.L. Hewett and  T.G. Rizzo ,Phys.Rev.Lett.{\bf 84},2080,(2000); K. Agashe, arXiv:0902.2400 [hep-ph]; K. Agashe, A. Azatov, L. Zhu, arXiv:0810.1016 [hep-ph]; K. Agashe, A. Falkowski, Ian Low, G.Servant, JHEP 
{\bf 0804}, 027,(2008).


\bibitem{davod2}H. Davoudiasl, J.L. Hewett and  T.G. Rizzo , Phys.Lett.{\bf B473} 43,(2000); K. Agashe {\it et al}, Phys.Rev.{\bf D76}, 115015,(2007); K. Agashe, S. Gopalakrishna, Tao Han, Gui-Yu Huang and A. Soni, arXiv:0810.1497 [hep-ph].  
\bibitem{our} .B. Mukhopadhyaya, S. Sen and S. SenGupta, 
Phys.Rev.Lett.{\bf 89}, 121101,(2002; Erratum-ibid.{\bf 89}, 259902,(2002);
Phys.Rev.{\bf D79}, 124029, (2009); Phys.Rev.{\bf D76}, 121501 (2007); 
Phys.Rev. {\bf D65} 124021,(2002). 
    
\bibitem{cvetic} K. Behrndt and M. Cvetic, Phys. Lett. {\bf B475} 253 (2000).


\bibitem{6dmodel1} A.~G.~Cohen and D.~B.~Kaplan,
  Phys.\ Lett.\  B {\bf 470}, 52 (1999);
  R.~Gregory,
  Phys.\ Rev.\ Lett.\  {\bf 84}, 2564 (2000);
  S.~M.~Carroll, S.~Hellerman and M.~Trodden,
  Phys.\ Rev.\  D {\bf 62}, 044049 (2000);
  N.~Arkani-Hamed, L.~J.~Hall, D.~Tucker-Smith and N.~Weiner,
  Phys.\ Rev.\  D {\bf 62}, 105002 (2000);
  Z.~Chacko and A.~E.~Nelson,
  Phys.\ Rev.\  D {\bf 62}, 085006 (2000);
  T.~Gherghetta and M.~E.~Shaposhnikov,
  Phys.\ Rev.\ Lett.\  {\bf 85}, 240 (2000).



\bibitem{6dmodel2} M.~Giovannini, H.~Meyer and M.~E.~Shaposhnikov,
  Nucl.\ Phys.\  B {\bf 619}, 615 (2001); P.~Kanti, R.~Madden and K.~A.~Olive,
  Phys.\ Rev.\  D {\bf 64}, 044021 (2001); C.~P.~Burgess, J.~M.~Cline, N.~R.~Constable and 
H.~Firouzjahi,
  JHEP {\bf 0201}, 014 (2002); M.~Giovannini,
  Phys.\ Rev.\  D {\bf 66}, 044016 (2002); M.~Giovannini,
  Class.\ Quant.\ Grav.\  {\bf 20}, 1063 (2003)

\bibitem{leblond} F. Leblond, R. C. Myers and D. J. Winters, JHEP
  {\bf{0107}} 031 (2001)

\bibitem{kogan} I. I. Kogan, S. Mouslopoulos, A. Papazoglou and
  G. G. Ross, Phys. Rev. {\bf D 64}, 124014 (2001)


\bibitem{dcssg}  D.~Choudhury and S.~SenGupta,
  Phys.\ Rev.\  D {\bf 76}, 064030 (2007)

\bibitem{rsj1} R.~Koley, J.~Mitra and S.~SenGupta,
  Phys.\ Rev.\  D {\bf 78}, 045005 (2008)
\bibitem{pol} See, for example, {\em String Theory}, J.Polchinski, 
Cambridge University Press., Cambridge, (1998).

\bibitem{lhc}A. A. Pankov, I. A. Serenkova, and A. V. Tsytrinov, Non. Pheno. Comp. Sys, {\bf 13}, 85, (2010); B. Laforge, hep-ph/0207166; P. Osland , A. A. Pankov, A. V. Tsytrinov and N. Paver, AIP Conf.Proc., {\bf 1149}, 219, (2009)
	


\end{thebibliography}
\end{document}